\documentclass[onecolumn,showpacs,amsmath,amssymb,prb]{revtex4}
\usepackage{graphicx}
\usepackage{dcolumn}
\usepackage{bm}

\begin{document}

\preprint{}

\title{Critical temperature of MgB$_2$ ultrathin superconducting films:\\ BCS model calculations in tight-binding approximation}

\author{Karol Sza\l{}owski}\email{kszalowski@wp.pl}
 \affiliation{Department of Solid State Physics, University of \L{}\'od\'z \\ul. Pomorska 149/153, PL 90-236 \L{}\'od\'z, Poland.}

\date{July 17, 2006}

\begin{abstract}
We develop the multi-band BCS model of superconductivity in the ultrathin films using the orthogonal tight-binding approximation for constructing the electron wavefunctions. This allows for relatively simple determination of the band structure near the Fermi level as well as the electron-electron interaction matrix elements of the BCS type. The model is applied to the ultrathin MgB$_2$(0001) films, for which the critical temperature values are calculated in the thickness range 2...10 MgB$_2$ layers. The importance of the boundary conditions is emphasised, as either boron or magnesium layers may cover the film. It is found that films thinner than 4 layers show substantial decrease in the critical temperature. The charge spillage outside the geometric boundaries of the film, which is allowed in our model, suppresses $T_c$ oscillations and weakens its dependence on film covering layer composition. \end{abstract}


\maketitle

\section{\label{sec:level1}Introduction}

The reduction of the superconducting film thickness leads to the occurrence of quantum size effects (QSE) which are caused by confining one of the electron co-ordinates in a quantum well. This leads to variation of such parameters as critical temperature and magnetic field, energy gap, chemical potential and heat capacity. The first analysis of such effects in the framework of Bardeen-Cooper-Schrieffer (BCS) theory of phonon-mediated superconductivity\cite{bcs}, based on free-electron model, was presented in the sixties \cite{blattthompson,thompsonblatt,paskinsingh}. The oscillatory behaviour of the  critical temperature $T_c$ and the chemical potential was predicted, together with the vital importance of boundary conditions for the solutions of the model. The problem of validity of common choices of boundary conditions was raised\cite{allen} and the modification\cite{yu} allowed for investigating a more realistic situation. The possible phonon quantization in the ultrathin film was also included in this model\cite{hwang}. Recently, the model was employed to determine the specific heat behaviour\cite{chen}.

Being aware of the possible limitations of the free-electron model, which does not involve neither crystalline nor band structure, in the present work we develop a tight-binding model for BCS-superconductivity. The basis for calculations is the multiband BCS model with bulk parameters adjusted to fit the experimental data. The aspects connected with the electronic structure of the ultrathin film are treated in the orthogonal tight binding approximation (TBA) that allows to find a simple relation between bulk and thin film matrix elements used in BCS model, as well as for determination of the necessary electronic densities of states and Fermi level shift.

As an example of application of the developed model, we present the calculations of the critical temperature and the energy gaps for magnesium diboride (MgB$_2$) films composed of a few monolayers.

According to the best of our knowledge, there has been no theoretical prediction of the superconducting parameters of ultrathin film published for this substance. The first-principle investigations of QSE in free-standing MgB$_2$ film by Huang \emph{et al.} concerned only the electronic structure and  film stability. Therefore we find it worthwhile to study the superconducting properties of such films within the framework of our model. We are convinced that magnesium diboride is the best candidate among the classical superconductors for applying TBA.    

The recently discovered magnesium diboride superconductor\cite{nagamatsu} is characterised by unexpectedly high critical temperature (39 K in bulk) while its chemical composition, crystalline structure as well as electron bands are rather uncomplicated. The discovery encouraged extensive and fruitful theoretical and experimental studies of various properties of this substance. The model of a phonon-mediated BCS-type superconductivity with two different energy gap values at different Fermi surface sheets was suggested\cite{choi,choiroundy} and its validity was experimentally proved by direct measurements of two gaps (see for example\cite{giubileo,chengap,szabo,tsuda1,tsuda2,iavarone,schmidt,gonnelli}) as well as heat capacity investigations\cite{bouquet}. The phonon-dependent mechanism of superconductivity in MgB$_2$ is supported by the studies of the isotope effect\cite{budko,hinks} and explained by means of the first-principle calculations\cite{yildirim}.

The theoretical studies of ultrathin magnesium diboride film properties are strongly motivated by the synthesis of good quality films of thickness down to single monolayers by Cepek \emph{et al.}\cite{cepek} via molecular beam epitaxy. The advantageous substrate for the growth of such films is (0001)Mg which possess the hexagonal surface symmetry and the lattice constants at the surface very close to that of MgB$_2$\cite{cepek}. However, to the best of our knowledge, no experimental report on the superconducting properties of ultrathin MgB$_2$ films has been published until now.

There exist some experimental evidence of QSE in the ultrathin superconducting films. The oscillatory changes of $T_c$ with Sn film thickness were found\cite{orr} in the eighties, however the effect was then attributed to QSE in the metallic grains making the film rather than in the film itself\cite{commentpaskin}. The latest experimental data are available mainly for the ultrathin Pb(111) films. The studies of Guo \emph{et al.}\cite{guo} detected the existence of superconductivity in 15-28 monolayers (ML) thick Pb films with the clear oscillations of the critical temperature, which was reduced towards its bulk value. In addition the variations of the electron-phonon coupling strength and the critical magnetic field were measured\cite{feng,bao} together with the variations of the normal-state resistance. The work of Eom \emph{et al.}\cite{eom} concerned Pb films of 5-18 ML thickness and both the decrease and the oscillatory behaviour of $T_c$ were confirmed. It is worth noticing that even in such ultrathin films consisting of just a few ML the superconducting properties are not suppressed by the existence of the fluctuations. The latest results show also the size-dependent reduction in $T_c$ and energy gap in the superconducting Nb grains (down to 8 nm in diameter)\cite{bose}.

\section{Multiband formulation of the BCS model}

The treatment of the problem is based on the BCS model\cite{bcs}, generalised to the case of multiple bands crossing the Fermi level, each of them assigned a different value of the energy gap. Such an extension of the BCS model was proposed by Suhl, Matthias and Walker\cite{suhl}. Its application is necessary for MgB$_2$ due to the Fermi surface structure. Moreover, in the case of the ultrathin film, every band additionally splits into two-dimensional subbands reflecting the electron confinement.  

Let us denote the band indices by $\alpha,\alpha'$ (including the band type as well as the discrete subband index which is valid for the thin film), while $\mathbf{k},\mathbf{k'}$ are the wavevectors form the first Brillouin zone (either three-dimensional or superficial). The reduced BCS hamiltonian of the grand canonical ensemble is then of the form \cite{bookbcs}:

\begin{equation}
H=\sum_{\alpha,\mathbf{k}}^{}{\epsilon_{\alpha\,\mathbf{k}}\,c^{\dagger}_{\alpha\mathbf{k}}c^{}_{\alpha\mathbf{k}}}+\!\!\sum_{\alpha,\mathbf{k},\alpha',\mathbf{k'}}^{}{\!\!\!\!V_{\alpha\,\mathbf{k},\alpha'\,\mathbf{k'}} \,c^{\dagger}_{\alpha'\,\mathbf{k'}}\, c^{\dagger}_{\alpha'-\mathbf{k'}} \,c_{\alpha-\mathbf{k}}\,c_{\alpha\mathbf{k}}}
\end{equation}

where $\epsilon_{\alpha\,\mathbf{k}}=E_{\alpha\,\mathbf{k}}-\mu$ is the energy referred to the Fermi level. This hamiltonian can be then diagonalized in the usual way\cite{bookbcs} in the mean field approximation. It must be emphasised that this approximation neglects the fluctuations of the order parameter and its validity has to be verified experimentally. We acquire the following self-consistent set of equations for the energy gap parameters $\Delta_{\alpha\mathbf{k}}(T)$: 
\begin{equation}
\Delta_{\alpha\mathbf{k}}=-\frac{1}{2}\sum_{\alpha,\mathbf{k'}}^{}{\frac{V_{\mathbf{k}\alpha,\mathbf{k'}\alpha'}\Delta_{\alpha'\mathbf{k'}}}{\sqrt{\epsilon_{\alpha'\mathbf{k'}}^2+\Delta_{\alpha'\mathbf{k'}}^2}}\tanh \frac{\sqrt{\epsilon_{\alpha'\mathbf{k'}}^2+\Delta_{\alpha'\mathbf{k'}}^2}}{2k_BT}}
\end{equation}

The isotropic \textit{s}-wave attractive interaction potential (in reciprocal space) is used in BCS approximation:
\begin{equation}
V_{\mathbf{k}\alpha,\mathbf{k'}\alpha'}=-V_{\alpha,\alpha'}\,\Theta\Bigl(E_D-\left|\epsilon_{\alpha\mathbf{k}}\right| \Bigr)\,\Theta\Bigl(E_D-\left|\epsilon_{\alpha'\mathbf{k'}}\right|\Bigr),
\end{equation}

with $V_{\alpha,\alpha'}>0$, i.e. the interaction is constant and nonzero only in the narrow shell of thickness $2E_D$ round each sheet of the Fermi surface. The energy cutoff $E_D=k_B\Theta_D \ll E_F$ is taken as a Debye energy ($\Theta_D$ being the Debye temperature) because the interaction is phonon-mediated. The choice of the interaction implies that also $\Delta_{\alpha\mathbf{k}}(T)$ may be nonzero only in this range and constant in each band in the vicinity of the Fermi surface. Assuming the constant electron single spin density of states (DOS) $g_{\alpha}(0)$ in the energy range of interest we get:
\begin{equation}
\Delta_{\alpha}(T)=\sum_{\alpha}^{}{2g_{\alpha'}(0)\,\Delta_{\alpha'}(T) V_{\alpha,\alpha'}\,F\bigl(\Delta_{\alpha'}(T),T\bigr)}
\end{equation}
where
\begin{equation}
F\left(\Delta_{\alpha}(T),T\right)\equiv\int_{0}^{k_B\Theta_D}{\!\!\frac{d\epsilon}{\sqrt{\epsilon^2+\Delta_{\alpha}^2(T)}}\,\tanh \frac{\sqrt{\epsilon^2+\Delta_{\alpha}^2(T)}}{2k_BT}}.
\end{equation}

The order parameter $\Delta_{\alpha}(T)$ is a decreasing function of the temperature and the critical temperature $T_c$ can be found by linearizing the set of equations (4) by setting $\Delta_{\alpha}(T_c)=0$, so that \begin{equation}
F(0,T_c)\equiv\int_{0}^{\Theta_D/T_c}{\frac{\tanh \frac{x}{2}}{x}\,dx}.
\end{equation}
Energy gap equations can be consequently written as:

\begin{equation}
\Delta_{\alpha} \Bigl(1-2g_{\alpha}(0)\, V_{\alpha,\alpha}\,F(0,T_c)\Bigr)-\sum_{\alpha'\neq\alpha}^{}{2\Delta_{\alpha'}g_{\alpha'}(0)\,V_{\alpha,\alpha'}\,F(0,T_c)}=0
\end{equation}
The above system is homogeneous at $\Delta_{\alpha}=0$ thus the solutions for $F(0,T_c)$ are the roots of the determinant of the characteristic matrix:
\begin{equation}
\label{eqn:tc} \det \Bigl(\delta_{\alpha,\alpha'}-2g_{\alpha}(0)\, V_{\alpha,\alpha'}\,F(0,T_c)\Bigr)=0.
\end{equation}

The smallest root corresponds to the highest temperature $T_c$ and has the physical importance as the transition temperature. Once $F(0,T_c)$ is known, the critical temperature is given by $T_c\simeq 1.134\,\Theta_D\,\exp(-F(0,T_c))$ if only $\Theta_D\gg T_c$. 

\section{Model of the thin film}

\subsection{Geometric model}

MgB$_2$ crystallizes in AlB$_2$ structure (hexagonal $\omega$ C32 structure) \cite{jones} with the lattice constants $a=$~3.083~\AA$\,$ and $c=$~3.521~\AA \cite{kong}. The crystal consists of the subsequent, equally separated parallel honeycomb graphene-like layers of B and hexagonal planes of Mg atoms. Owing to this layered structure, the typical orientation of ultrathin films is (0001) (hexagonal axis-oriented) and we limit ourselves to considering only such films. The $z$-direction is therefore chosen parallel to the hexagonal axis. The single boron plane consists of two sublattices. Each boron atom has two nearest neighbours (n.n.) at the distance of $c$ (in the direction perpendicular to the plane) in the same sublattice and 3 n.n. at the distance of $a/\sqrt{3}$ (in-plane) in the other sublattice. 

For the thin film, either B or Mg atoms can cover the film from each side. Three possible configurations exist which we denote by B...B, Mg...Mg and Mg...B.       

\begin{figure}
\includegraphics[scale=0.7]{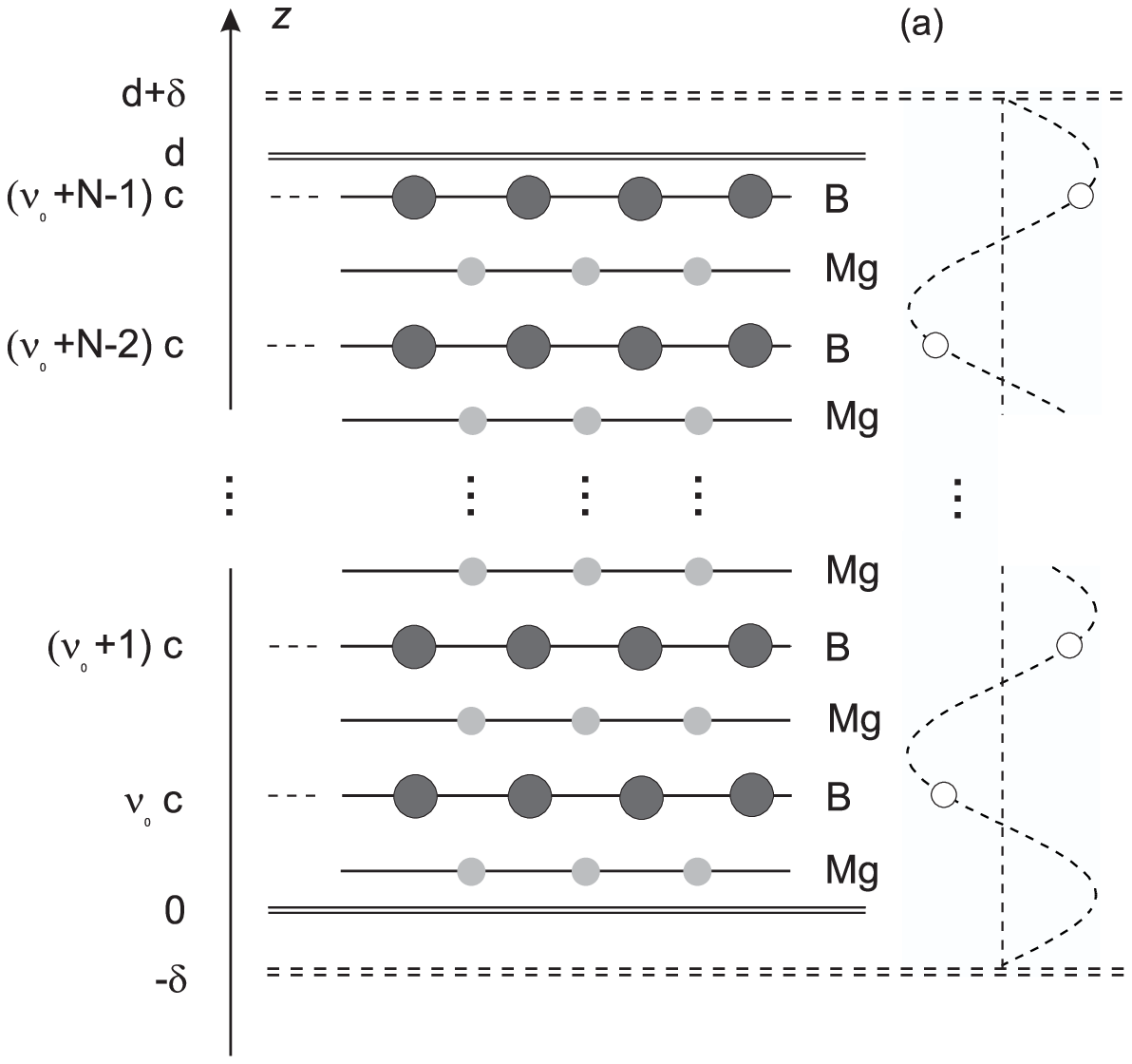}
\includegraphics[scale=0.4]{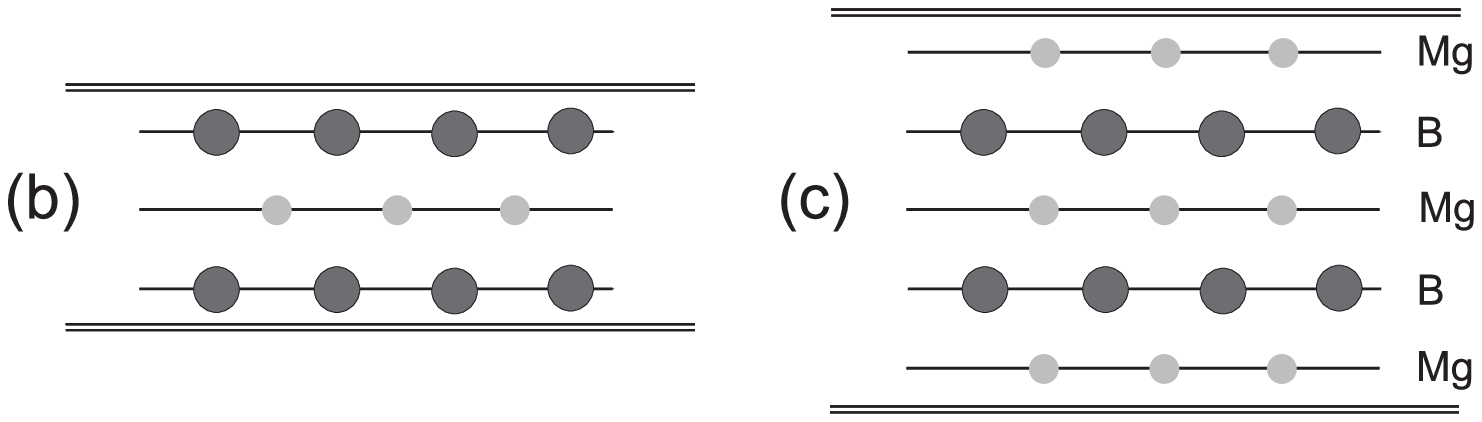}
\caption{\label{fig:filmmodel}(a) Schematic side view of the (0001)-axis oriented thin MgB$_2$ film of Mg...B covering type, containing $N$ boron layers. Solid double lines are the geometric boundaries of the film while dashed ones are the boundaries of the assumed infinite quantum well for the electrons in $z$ direction. The dashed curve on the right depicts the behaviour of $\Gamma$ amplitude. (b) The view of the B...B type film (c) The view of the Mg...Mg type film, both for $N$=2.}
\end{figure}

As we explain further, the band structure calculations in TBA require only boron electronic orbitals, so that only the boron atoms positions inside the film are important. The position vector of B atoms in the $z$ direction (perpendicular to the film plane) can be written as $\mathbf{R_{\nu}}=\left(\nu+\nu_0\right)\mathbf{c}$ where $\nu=0,\dots,N-1$ numbers the subsequent boron layers. The $\nu_0$ parameter describes the position of the first boron plane towards the geometric boundary of the film and depends on the layer which covers the film from the substrate side: $\nu_0=1/4$ for B layer while $\nu_0=3/4$ for Mg layer. It is assumed that the relaxation of the atomic layers in the vicinity of surface does not lead to noticeable deviation of interplanar distances from the bulk value\cite{li}.

The full MgB$_2$ layer consists of one B plane and one Mg plane and has the thickness $c$ thus the total thickness of the film is $d=N_{MgB_2}c$ where $N_{MgB_2}$ is a number of such layers. For the Mg...B film consisting of $N$ boron layers we have $N_{MgB_2}=N$ while the fractional values are obtained in the other situations: $N_{MgB_2}=N-1/2$ for the B...B film and $N_{MgB_2}=N+1/2$ for the Mg...Mg film. This implies also that the B...B film with $N$ boron layers is assigned the same thickness as the Mg...Mg film containing $N+1$ boron layers. 

\subsection{Electron wavefunction in the orthogonal TBA}

The normalized basis bulk electron wavefunctions $\psi_{\alpha\mathbf{k}}$ in orthogonal TBA\cite{callaway} are constructed as follows: 
\begin{equation}
\psi_{\alpha\mathbf{k}}\left(\mathbf{r}\right)=\frac{1}{\sqrt{\mathcal{N}}}\sum_{\mathbf{R}}^{}{e^{i\mathbf{k}\mathbf{R}}\,\varphi\left(\mathbf{r}-\mathbf{R}\right)}.
\end{equation}

Functions $\varphi$, denoted then as 'atomic orbitals', may in general be the superpositions of the orbitals possessing different symmetry or centred on the neighbouring sites of the nonequivalent sublattices, as it is for MgB$_2$.  $\mathbf{R}$ vectors denote the positions of $\varphi$ orbital centres.

For the ultrathin film case, the translational symmetry in $z$ direction is broken and the nearest neighbour number for the atoms inside the superficial atomic monolayers is decreased which requires modification of the method. The electrons are confined in a potential well in the $z$ direction. Usually the infinite square well model is used in spite of the fact that the actual well is of finite depth. The well width may equal the geometrical thickness of the film $d$ implying that the electron wavefunctions vanish at the geometrical boundaries (this is the case of boundary conditions considered by Thompson and Blatt\cite{thompsonblatt,blattthompson}). 

In a more realistic situation, the electron density spillage beyond the geometric boundaries over a small distance $\delta$ is allowed so that the width of the well is $d+2\delta$. Such a generalization of the model was proposed by Yu \emph{et al.}\cite{yu} and the parameter $\delta$ was selected to obtain the bulk-like charge density exactly in the middle of the film. If the distance $\delta$ is energy-dependent, the model can be made equivalent to the phase-accummulation model in which the phase of a wavefunction changes as a result of reflections from the boundaries (as introduced by Echenique and Pendry\cite{echenique}). However, in the present paper the spillage distance $\delta$ is taken constant. 

In the original formulation for the free-electron model, the spillage distance $\delta$ was determined uniquely by the bulk Fermi wavevector by demanding that the charge density in the geometric middle of the film equals its bulk value. Such a procedure implied that the Fermi wavevector and Fermi energy for a thin film were unchanged with respect to the bulk situation. The key assumption of local charge neutrality in the middle of the film was criticised by Rogers \emph{et al.}\cite{rogers}. 

In our work the parameter $\delta$ is treated as a free parameter which may be adjusted to fit best the experimental data for the specified substrate and covering of the real film. Especially, we do not require the Fermi level to be thickness-independent. On the contrary, we allow it to shift to guarantee  the bulk-like electronic density averaged over the whole film thickness.

 The extended infinite potential well method is commonly applied to investigate the ultrathin film properties (for example see the studies of Pb film characteristics in the free-electron approximation\cite{czoschke3,czoschke1,czoschke2}).

The wavevectors are $\mathbf{k}=\mathbf{k^{\|}}+\mathbf{k^{\perp}}$, where $k_z=\left|\mathbf{k^{\perp}}\right|$. In anticipation of the appearence of discrete $k_z$ values, we replace the index $\mathbf{k}$ with $\mathbf{k^{\|}}$ and $n$. Our construction of the trial electron wavefunctions follows the calculation of Szczeniowski and Wojtczak \cite{wojtczak} performed to characterize the magnetic properties of the thin films within the band model. The factor $\exp\left(ik_z z\right)$ satisfying the Bloch condition in $z$ direction for the bulk crystal is replaced with the more general $z$-dependent amplitude \cite{wojtczak} $\Gamma\left(k_z;\mathbf{R_{\nu}}\right)$. The wavefunction is then in the form: 

\begin{equation}
\psi_{\alpha\mathbf{k^{\|}}n}\left(\mathbf{r}\right)\!=\!\frac{1}{\sqrt{\mathcal{N}}}\!\sum_{\mathbf{R^{\|}},\nu}^{}{\!\Gamma\!\left(n;\mathbf{R_{\nu}}\right)e^{i\mathbf{k^{\|}}\mathbf{R^{\|}}}\varphi\!\left(\mathbf{r}-\mathbf{R^{\|}}\!-\mathbf{R^{\perp}_{\nu}}\!\right)}
\end{equation}

In order to have an orthonormal set of wavefunctions, the following condition:
\begin{equation}\displaystyle\sum_{\nu=0}^{N-1}{\Gamma^*\left(n';\mathbf{R_{\nu}}\right)\Gamma\left(n;\mathbf{R_{\nu}}\right)}=\delta_{n'n}
\end{equation}
must be satisfied.

Taking in consideration the geometry of the ultrathin film we impose the following boundary conditions on the amplitudes which vanish at the assumed boundaries of the potential well in $z$ direction, as shown in Fig. \ref{fig:filmmodel}.
\begin{equation}
\Gamma\left(n;-\delta\right)=0\quad \Gamma\left(n;d+\delta\right)=0
\end{equation}

The amplitudes are given by:
\begin{equation}
\Gamma\left(n;\mathbf{R_{\nu}}\right)=C(n)\sin\Bigl[\left(\nu+\nu_0\right)ck_z^n+\delta k_z^n\Bigr]
\end{equation}

with the normalization constant:
\begin{equation}
C(n)=\left(\,\sum_{\nu=0}^{N-1}{\sin^2\Bigl[\left(\nu+\nu_0\right)ck^z_n+\delta k_z^n\Bigr]}\right)^{-1/2}
\end{equation}
since the atomic orbitals are assumed to be orthonormal: $\int_{}^{}{\varphi^*\left(\mathbf{r}-\mathbf{R'}\right)\varphi\left(\mathbf{r}-\mathbf{R}\right)\,d\,^3r}=\delta_{\mathbf{R}-\mathbf{R'}}$. In the case of no electron density spillage outside the geometric boundaries ($\delta=0$) we have $C(n)=\sqrt{2/N}$.
  
In the potential well there exist $N$ allowed discrete wavevector values, describing the quantum-well states:
\begin{equation}
k_z^n=\frac{n\pi}{d+2\delta},
\end{equation}
for $n=1,\dots,N$.

\subsection{Energy bands and Fermi level}

The Fermi surface of MgB$_2$ is created by two $\pi$-type and two $\sigma$-type electron bands\cite{kortus}, first of whom form 2D honeycomb-like network while the second are cillinder-shaped and possess 3D character. In the vicinity of the Fermi surface, orthogonal TBA method provides the reasonable description of energy bands\cite{kortus,kong,an}. 

The $\pi$ band origins from hybridization of $p_z$ orbitals centred at boron atoms of two B sublattices\cite{kong}. Orthogonal TBA procedure with respect to such orbitals leads to the hamiltonian matrix
\begin{equation}
H_{\pi}=\left(\begin{array}{cc}e_{\pi}\!+2t_{\perp}\cos ck_z & t'_{\|}\left
(2e^{-i\frac{ak_x}{2\sqrt{3}}}\cos\frac{ak_y}{2}+e^{i\frac{ak_x}{\sqrt{3}}}\right)\\ & \\
h.c. & e_{\pi}\!+2t_{\perp}\cos ck_z\\\end{array}\right)
\end{equation}

and then to the dispersion relations:
\begin{align}
\label{eqn:dispersionpi}
&\epsilon_{\pi^{\pm}}(\mathbf{k})=e_{\pi}+2t_{\perp}\cos ck_z \nonumber \\
&\pm t'_{\|}\sqrt{1+4\cos\frac{ak_y}{2}\left(\cos\frac{ak_y}{2}+\cos\frac{ak_x\sqrt{3}}{2}\right)}
\end{align}

describing the $\pi$ bonding and antibonding band. After Kong \emph{et al.}\cite{kong} we accept the parameter values $e_{\pi}=$~0.04~eV, $t_{\perp}=$~ 0.92~eV and $t'_{\|}=$~1.60~eV (the Fermi level is set to 0).

The same procedure for the thin film affects only the diagonal hamiltonian elements describing hopping terms between n.n. in the same sublattice (which lie in the layers above and below the given B atom). The non-diagonal elements remain unaltered since the n.n. from the second sublattice lie in plane with the specified atom. In addition, an extra diagonal term $E_F-E_F^{\infty}$ appears to allow for chemical potential change to preserve the constant electron density averaged over the film thickness. This causes the dispersion relation (\ref{eqn:dispersionpi}) to be modified in the following way:

\begin{eqnarray} 
\label{eqn:diagonal}2t_z\cos ck_z&\rightarrow& E_F-E_F^{\infty}+\nonumber\\  &+&\Gamma(n;\mathbf{R_0})\Gamma(n;\mathbf{R_1})+\Gamma(n;\mathbf{R_{N-2}})\Gamma(n;\mathbf{R_{N-1}}) \nonumber \\
&+&\sum_{\nu=1}^{N-2}{\Gamma\left(n;\mathbf{R_{\nu}}\right)\Bigl(\Gamma\left(n;\mathbf{R_{\nu+1}}\right)+\Gamma\left(n;\mathbf{R_{\nu-1}}\right)\Bigr)}.\nonumber\\
\end{eqnarray}

The $\sigma$ bands are formed by overlapping of the two-center bonding $sp^2$ orbitals, centered between the given atom and its three n.n. from the other sublattice\cite{kong}. The bulk hamiltonian is:
\begin{widetext}
\begin{align}
H_{\sigma}&=e_{\sigma}+2t_{\perp}\cos ck_z\,\left(
\begin{array}{ccc}
1 & 0 & 0\\ 0 & 1 & 0 \\ 0 & 0 & 1
\end{array}\right)+\nonumber \\
\\
&+\left(
\begin{array}{ccccc}
0 &\quad & 2t'_{\|}\cos\frac{a\left(k_x\sqrt{3}-k_y\right)}{4}+2t''_{\|}\cos\frac{a\left(k_x\sqrt{3}+3k_y\right)}{4} & \qquad\qquad & 2t'_{\|}\cos\frac{a\left(k_x\sqrt{3}+k_y\right)}{4}+2t''_{\|}\cos\frac{a\left(-k_x\sqrt{3}+3k_y\right)}{4}\\ & \qquad\qquad & & \qquad\qquad & \\
h.c. & \qquad\qquad & 0 & \qquad\qquad & 2t'_{\|}\cos\frac{ak_y}{2}+2t''_{\|}\cos\frac{ak_x\sqrt{3}}{2}\\ & \qquad\qquad & &\qquad\qquad & \\
h.c. &\qquad\qquad & h.c. & \qquad\qquad & 0
\end{array}\right),
\end{align}
\end{widetext}

with the parameters\cite{kong} $e_{\sigma}=$~-12.62~eV, $t_{\perp}=$~0.094~eV, $t'_{\|}=$~5.69~eV and $t''_{\|}=$~0.91~eV.

The two bands (heavy and light holes) crossing the Fermi level are, after expanding in the vicinity of $\Gamma A$ line in the Brillouin zone:
\begin{eqnarray}
\epsilon_{\sigma}(\mathbf{k})&=&e_{\sigma}+2\left(t'_{\|}+t''_{\|}\right)-2t_{\perp}\cos ck_z -\left(k_x^2+k_y^2\right)\frac{t'_{\|}a^2}{4} \nonumber \\
\epsilon_{\sigma}(\mathbf{k})&=&e_{\sigma}+2\left(t'_{\|}+t''_{\|}\right)-2t_{\perp}\cos ck_z -\left(k_x^2+k_y^2\right)\frac{3t''_{\|}a^2}{4}. \nonumber \\
\end{eqnarray}

The modification due to the ultrathin film geometry is fully analogous to that carried out for $\pi$ bands. The diagonal element and thus the dispersion relation acquires the terms (\ref{eqn:diagonal}) instead of $2t_{\perp}\cos ck_z$. 

The chemical potential $\mu$ is related to the electron density $n_e$ in the standard way, i.e. at $T=0$ we have the Fermi energy $\mu=E_F$ and $n_e=\sum_{\alpha,\mathbf{k}}^{}{\Theta\left(E_{\alpha\mathbf{k}}-E_F\right)}$. This leads to the formulas:
\begin{eqnarray}
\label{eqn:efbulk} n_e&=&\sum_{\alpha}^{}{\int_{0}^{E_F}{\!\!g_{\alpha}(E)\,dE}} \\
\label{eqn:eftf} Nn_e&=&\sum_{\alpha}^{}{\sum_{n=1}^{N}{\int_{0}^{E_F}{\!\!g_{\alpha,n}(E)\,dE}}},
\end{eqnarray}
where the first one is for the bulk case (the summation over $\mathbf{k}$ is over the 3D Brillouin zone and the electron density is per 3D unit cell) while in ultrathin film we obtain the second relation (the summation over $\mathbf{k}$ is over the surface Brillouin zone and the electron density per surface unit cell). The DOS for each band can be calculated numerically from the dispersion relation.

\subsection{Interaction potential matrix elements}
The matrix element of the interaction potential in the reciprocal space between $\alpha\mathbf{k}$ and $\alpha'\mathbf{k'}$ electrons is:
\begin{equation}
\mathcal{V}_{\alpha'\mathbf{k'},\alpha\mathbf{k}}=\int_{}^{}{\!\!\!\!\int_{}^{}{\left|\psi_{\alpha'\mathbf{k'}}\right|^2 V\left(\mathbf{r},\mathbf{r'}\right)\left|\psi_{\alpha\mathbf{k}}\right|^2\,d^3 r \,d^3 r'}}.
\end{equation}

The potential in configurational space $V\left(\mathbf{r},\mathbf{r'}\right)$ is the potential of the screened Coulomb interaction and if the screening is strong enough, we approximate it by the contact potential $V_{\alpha'\mathbf{k'},\alpha\mathbf{k}}\,\delta^3\left(\mathbf{r}-\mathbf{r'}\right)$ (and such an approximation was used in the free-electron model studies\cite{thompsonblatt,blattthompson,paskinsingh}). 

In the spirit of the orthogonal TBA, we assume that the nonvanishing contribution to the integral comes only from the electron scattering between the atomic orbitals centered on the same lattice site, i.e. from the terms proportional to $\left|\varphi_{a'}\left(\mathbf{r}-\mathbf{R}\right)\right|^2\left|\varphi_{a}\left(\mathbf{r}-\mathbf{R}\right)\right|^2$, where $a'$ and $a$ denote the kind of atomic orbitals used to construct the wavefunctions $\psi_{\alpha'\mathbf{k'}}$ and $\psi_{\alpha\mathbf{k}}$, respectively. 

This assumption allow us to express both the bulk and ultrathin film matrix elements as:
\begin{widetext}
\begin{eqnarray}
\mathcal{V}^{bulk}_{\alpha'\mathbf{k'},\alpha\mathbf{k}}&=&\frac{V_{\alpha'\mathbf{k'},\alpha\mathbf{k}}}{\mathcal{N}^2}\sum_{\mathbf{R}}^{}{\int_{}^{}{\left|\varphi_{a'}\left(\mathbf{r}-\mathbf{R}\right)\right|^2\left|\varphi_{a}\left(\mathbf{r}-\mathbf{R}\right)\right|^2 d^3 r}}=\frac{V_{\alpha',\alpha}}{\mathcal{N}}\Theta\bigl(E_D-\left|\epsilon_{\alpha\mathbf{k}}\right| \bigr)\Theta\bigl(E_D-\left|\epsilon_{\alpha'\mathbf{k'}}\right|\bigr)\!\!\int_{}^{}{\left|\varphi_{a'}\left(\mathbf{r}\right)\right|^2 \left|\varphi_{a}\left(\mathbf{r}\right)\right|^2 d^3 r}\nonumber\\
\mathcal{V}_{\alpha'n',\alpha n}&=&\frac{V_{\alpha',\alpha}}{\mathcal{N}}\Theta\bigl(E_D-\left|\epsilon_{\alpha\mathbf{k}}\right| \bigr)\Theta\bigl(E_D-\left|\epsilon_{\alpha'\mathbf{k'}}\right|\bigr) \sum_{\nu=0}^{N-1}{\Gamma^2 (n';\mathbf{R_{\nu}})\,\Gamma^2 (n;\mathbf{R_{\nu}}) \!\!\int_{}^{}{\left|\varphi_{a'}\left(\mathbf{r}\right)\right|^2\left|\varphi_{a}\left(\mathbf{r}\right)\right|^2 d^3 r}}
\end{eqnarray}
\end{widetext}

It is evident that in the orthogonal TBA the thin film matrix element is expressed only by its bulk value (for appropriate bulk bands $\alpha'$, $\alpha$) and the purely geometric factor (dependent on discrete subband indices $n'$ and $n$):

\begin{equation}
\label{eqn:matrix} \mathcal{V}_{\alpha'n',\alpha n}=\mathcal{V}^{\,bulk}_{\alpha',\alpha}\,\sum_{\nu=0}^{N-1}{\,\Gamma^2(n';\mathbf{R_{\nu}})\,\Gamma^2(n;\mathbf{R_{\nu}})} 
\end{equation}

The presented method does not take into account the quantization of phonons due to the ultrathin film geometry. In our opinion, this simplification seems justified for the case of MgB$_2$. The studies of Yildirim \emph{et al.}\cite{yildirim} shows that the main contribution to the creation of the superconducting state results from the interaction of electrons with the E$_{2g}$ mode optical phonons causing in-plane deformation of the honeycomb B layers. These modes should remain unmodified in (0001)-oriented ultrathin film since their displacement vectors lie in this plane. The possible surface phonon modes are also neglected.  

We do not include the surface electronic states in our study and neglect the possible modification of the electron-phonon coupling at the surface, in analogy to the free-electron studies. In a recent paper Petaccia \emph{et al.}\cite{petaccia} measured the electron-phonon coupling for the surface state in MgB$_2$. The presence of a surface state was also detected for 18~ML MgB$_2$ film by the same authors. The contribution of the electronic surface states to the superconductivity of bulk MgB$_2$ was detected by Souma \emph{et al.}\cite{souma}. On the other hand, the calculations of band structure for ultrathin films by Huang \emph{et al.}\cite{huang} predicted the presence of a surface state at least 0.3 eV above the Fermi level (for the thinnest film; the state shifts upwards in energy when the film becomes thicker), so that it should not participate in superconductivity.

\section{Results}

We perform the calculations for the films composed of 2 to 10 MgB$_2$ layers, corresponding to the thickness range 7.0~\AA to 35.2~\AA. For each number of the boron monolayers we consider all the possible compositions (i.e. the B...B, Mg...Mg and Mg...B covered structures). Apart from the typical choice of the charge spillage distance $\delta$~=0, we also use the value of 0.5~\AA.

The electronic DOS is calculated from the known dispersion relations using the Monte Carlo technique. A number of $10^8$ random wavevectors from the approprate first Brilloiun zone (3D for bulk or 2D for thin film) is generated and the energy values are computed. 

To determine the Fermi level shift, the single spin DOS is calculated for energy intervals of 0.05~eV width. For bulk crystal case, the electron density is obtained by evaluating the integral (\ref{eqn:efbulk}) with the result $n_e=$~2.9. The occupancies of the individual subbands are: 0.97 ($\sigma$, light holes), 0.94 ($\sigma$, heavy holes), 0.93 (bonding $\pi$) and 0.06 (antibonding $\pi$). This properly describes the character of each subband (hole-like or electron-like)\cite{kortus}. For ultrathin film, the integral (\ref{eqn:eftf}) (with unaltered $E_F$) is performed first and the $E_F$ shift necessary to preserve $n_e$ is computed by the linear interpolation on the basis of the values of (\ref{eqn:eftf}) tabulated in 0.05~eV intervals.

\begin{figure}
\includegraphics[scale=0.7]{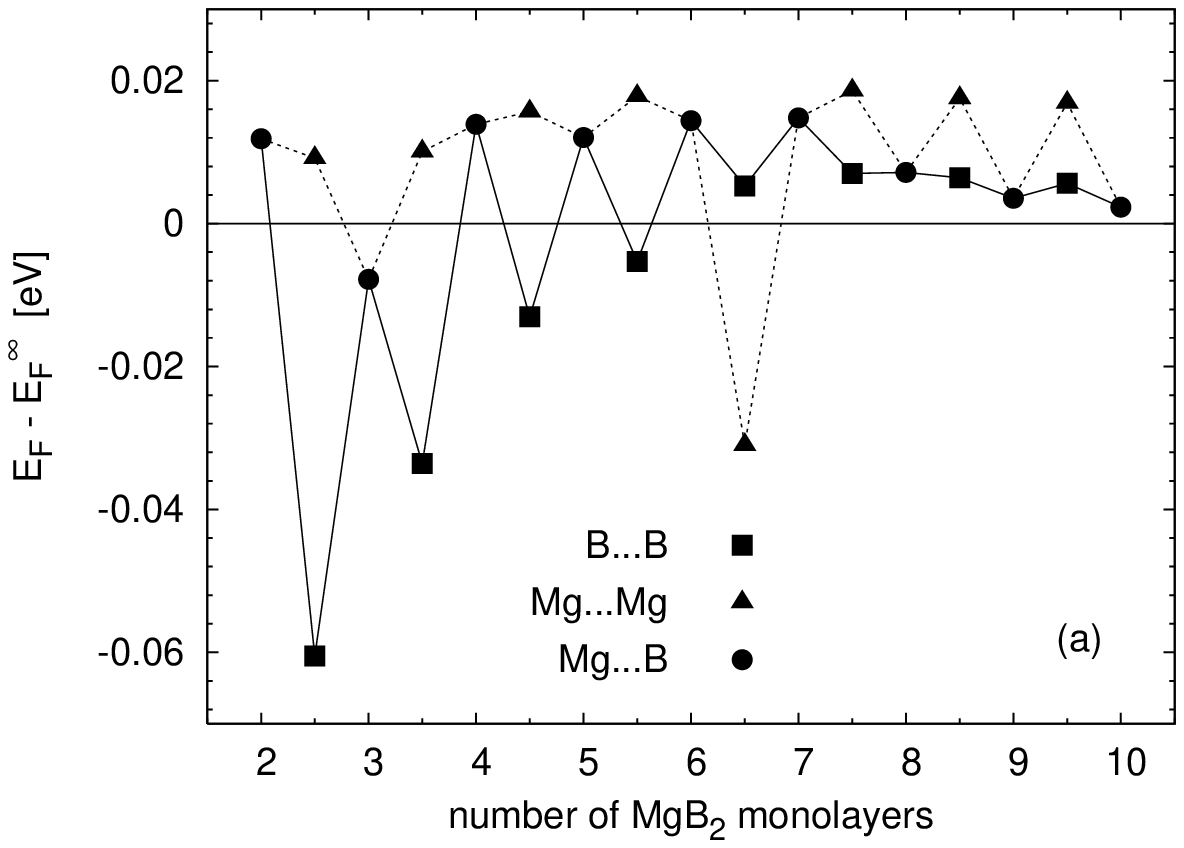}
\includegraphics[scale=0.7]{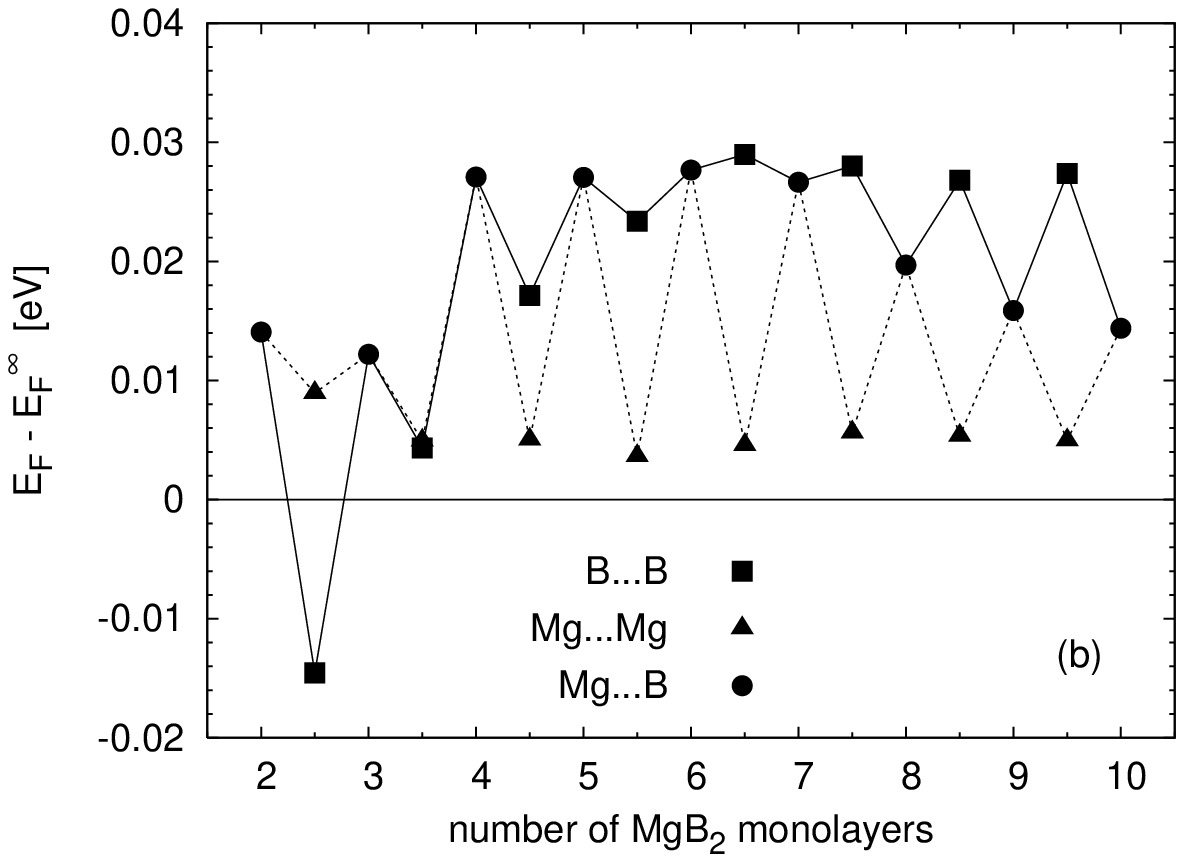}
\caption{\label{fig:fermilevel}The Fermi level change computed in the tight binding model for MgB$_2$ ultrathin film, for $\delta=$~0.0~\AA  (a) and $\delta=$~0.5~\AA  (b).}
\end{figure}

The obtained Fermi level shift with respect to the bulk value $E_F^{\infty}$ (which equals 0 for the parametrization of Kong \emph{et al.}) is presented in Fig. \ref{fig:fermilevel}. It is visible that the deviation does not exceed 0.06~eV (with no charge spillage) while it is even less for the charge spillage allowed. The clear short-period oscillatory changes with film thickness are present when switching between different compositions of the covering layers. Following the variability of $E_F$ within fixed covering layers, we observe more monotonic behaviour. 

It is worth noticing that the free-electron model with the boundary conditions of Thompson and Blatt\cite{thompsonblatt,blattthompson} predicted the increase of the Fermi energy when reducing the film thickness, while the opposite was true for the considerations performed by Paskin and Singh\cite{paskinsingh}. In both studies the oscillatory pattern was rather weak. In the original formulation of the model with charge spillage by Yu \emph{et al.}\cite{yu}, the Fermi level of a thin film was kept approximately fixed as a consequence of the specific choice of the spillage distance $\delta$.

The electronic DOS in the vicinity of $E_F$ is calculated analogously by MC method for the interval width 0.18~eV (the thickness of the energy range in which BCS interaction is nonzero).     
  
For the bulk case we obtain the total DOS of $g_{\pi}=$~0.28~eV$^{-1}$ and $g_{\sigma}=$~0.15~eV$^{-1}$ for the given tight-binding parametrization\cite{kong}. This may be compared with $g_{\pi}=$~0.204~eV$^{-1}$ and $g_{\sigma}=$ ~0.150~eV$^{-1}$ resulting from the LDA-based calculations\cite{liu}. It is visible that the DOS in $\sigma$ bands is reproduced exactly but the TBA significantly overestimates the DOS for $\pi$ bands.

The bulk matrix elements for the BCS interaction are determined such that they lead to the consistency with the experimental $T=0$ energy gap values as well as critical temperature as taken from the measurements of Gonnelli \emph{et al.} \cite{gonnelli} for monocrystals: $\Delta_{\pi}(0)=$~2.80~meV, $\Delta_{\sigma}(0)=$~7.1~meV and $T_c=$~37.6~K. The Debye temperature is $\Theta_D=$~1050~K \cite{bouquet} (see the further disussion). We obtain the following matrix elements: $V_{\sigma,\sigma}=$~0.694, $V_{\sigma,\pi}=$~0.353 and $V_{\pi,\pi}=$~0.056.

\begin{figure}
\includegraphics[scale=0.7]{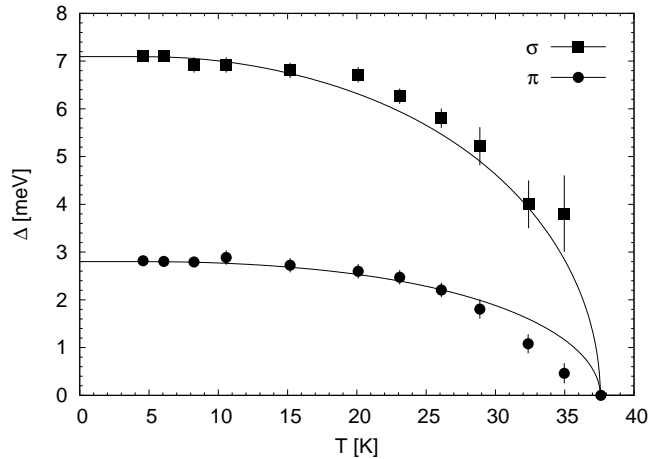}
\caption{\label{fig:deltabulk}Temperature dependence of bulk energy gap values. The experimental values are taken from \cite{gonnelli}. The solid line is a result of our calculations in the BCS model with the TBA-obtained DOS.}
\end{figure}

The usefulness of the BCS model with such parametrization is tested by comparison with the experimental data for bulk energy gaps dependence on temperature, presented in Fig. \ref{fig:deltabulk}.  

\begin{figure}
\includegraphics[scale=0.7]{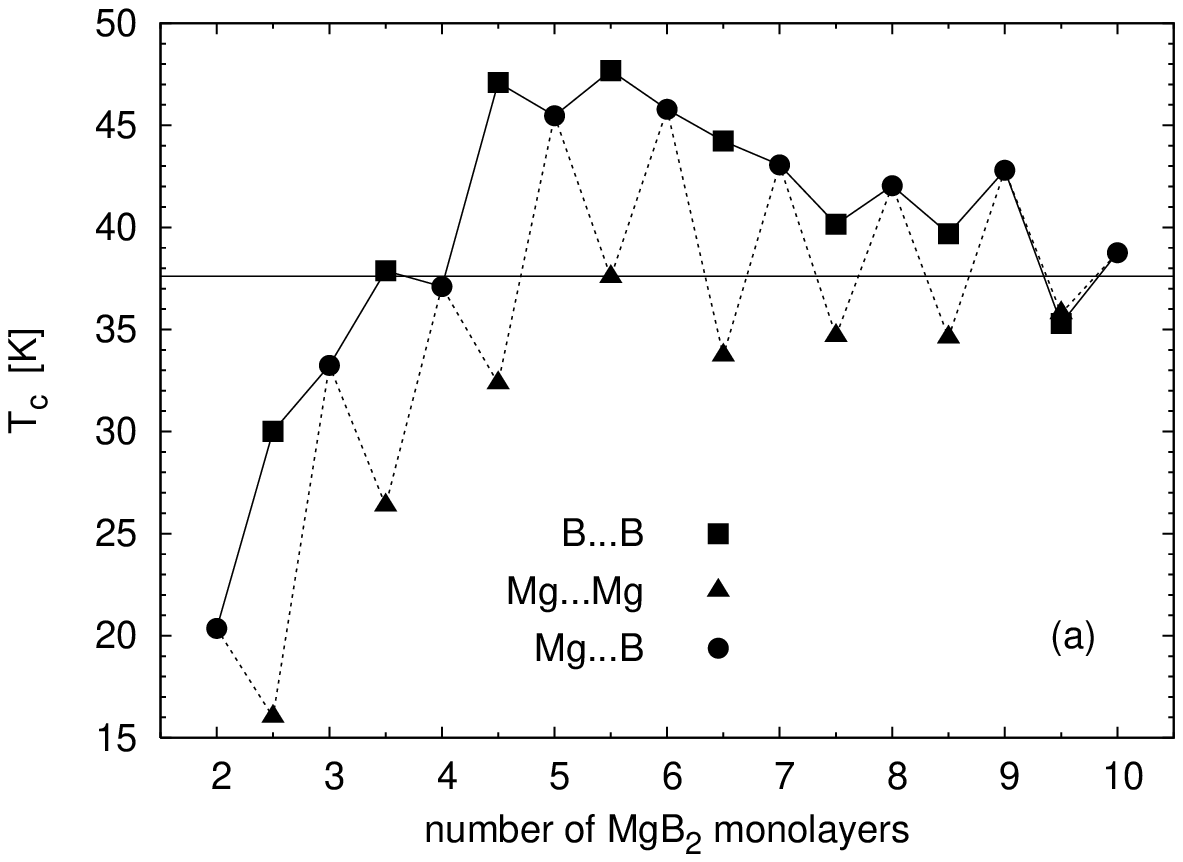}
\includegraphics[scale=0.7]{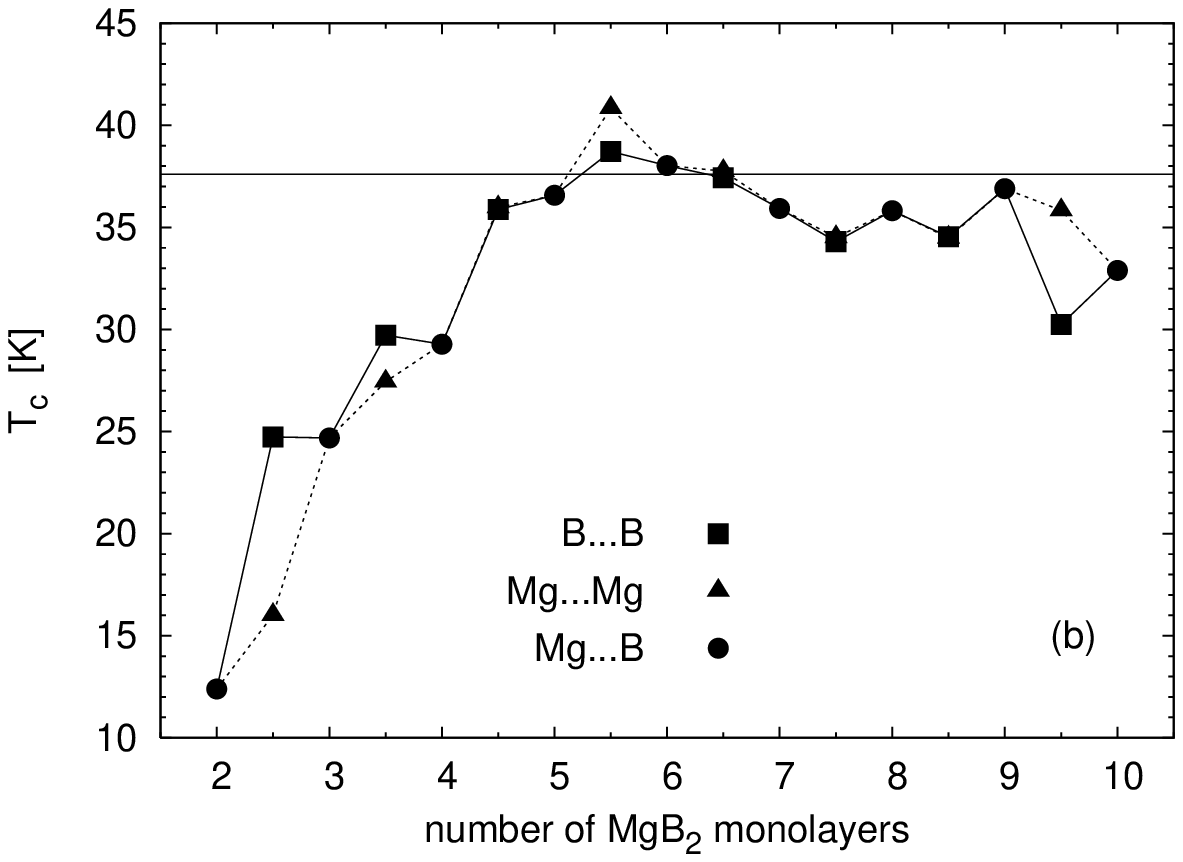}
\caption{\label{fig:tc}Critical temperature of MgB$_2$ ultrathin films calculated according to the model developed, for $\delta=$~0.0~\AA  (a) and $\delta=$~0.5~\AA  (b). The dashed line is bulk value.}
\end{figure}

The excellent agreement with the $\sigma$ energy gap values is visible while the $\pi$ energy gap behaviour is reproduced well by the two-band BCS model apart from the range close to critical temperature. 

As to the shape of $ \Delta_{\pi}\left(T\right) $ function, it appears to us that this is the matter of choice of the cutoff energy $k_B\Theta_D$. The cutoff parameter is commonly set to be a Debye energy or at least of this order(see\cite{parks}). It is visible in Fig.~2 in the paper of Liu \emph{et al.}\cite{liu}, presenting the BCS-based calculations, that the ratio $T_c/\Theta_D=$~0.46. For the critical temperature of 39 K we get the cutoff temperature $\Theta_D=$~85~K ($k_B\Theta_D=$~7.3~meV). The measured Debye temperature ranges from 750~K\cite{budko} to 1050~K\cite{bouquet}. It is also visible from the visual inspection of Fig.~1\cite{liu} (energy dependence of the Eliashberg function) that the phonons involved mostly in the superconducting state have the energies of the range 60~meV, which is an order of magnitude larger than the cutoff energy needed to depict accurately the flattening of the curve. 
It is possible to reproduce better the $\Delta_{\pi}\left(T\right)$ shape in two-band model at the expense of using very low cutoff energy. However, we decided to select the cutoff consistent with the experimental Debye temperature data.

\begin{figure}
\includegraphics[scale=0.7]{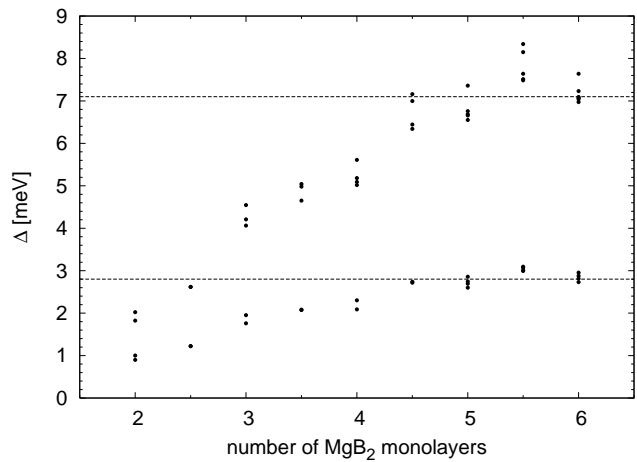}
\caption{\label{fig:gaps}Superconducting gap parameters at $T=$~0 for MgB$_2$ ultrathin films, $\delta=$~0.5~\AA. The integer thicknesses correspond to Mg...B covering, while half-integer are for Mg...Mg structure. The B...B type covering is omitted for clarity. The dashed lines are the bulk gaps for $\pi$ and $\sigma$ band, according to\cite{gonnelli}.}
\end{figure}

In the ultrathin film case, the 2D DOS in $\sigma$ band equals the 3D DOS due to negligible dispersion in the $z$ direction (i.e. the Fermi surface sheet is almost exactly cylindric). 

The critical temperatures for the ultrathin films were determined from the equation (\ref{eqn:tc}), with the matrix elements of the form given by (\ref{eqn:matrix}). Due to the existence of $N$ wavevectors $k_z$ for the film composed of $N$ boron planes, a number of $N$ gaps may exist for each band $\sigma$ or $\pi$. As the DOS in the $\sigma$ band is independent on the $n$ value, an energy gap is assinged to each of $N$ subbands. In the case of the $\pi$ band, not every $n$-th subband crosses the Fermi level so that the number of gap values is less than $N$.   

The calculated critical temperatures are presented in Fig.~\ref{fig:tc}. The short-period oscillatory character of changes can be observed with substantial decrease of $T_c$ for the films thinner than 4 layers. The oscillations, sharp for $\delta=$~0, reflect the changes of boundary conditions for different film coverings. The effect of nonzero charge spillage on $T_c$ may be compared with the free-electron model calculations of Yu \emph{et al.}\cite{yu}. In general, the charge spillage leads to decrease of critical temperature below its bulk value (which is consistent with the behaviour observed experimentally and thus justifies the necessity of including charge spillage in the model). Without charge spillage, $T_c$ is increased (both for the boundary conditions used by Thompson and Blatt\cite{thompsonblatt,blattthompson} and Paskin and Singh\cite{paskinsingh}), an effect which is not confirmed by the available experimental data. It is worth noticing that the oscillations with film thickness are strongly supressed when charge spillage takes place. In this situation also the chemical composition of the covering layers (B or Mg) has only weak effect on the $T_c$, contrary to the situation for $\delta=$~0. Unlike the free-electron model, which predicts the oscillation period $\pi/k_F$, the presented model does not give such a clear result, related to the Fermi wavevector, which can be attributed to the rather complicated Fermi surface geometry. 

Another remark concerning the comparison of our results with the free-electron based ones appears important. In such studies the film thickness is treated as a continuous parameter. For any real crystalline structure, we obviously obtain the discrete thicknesses and the character of the plot of any parameter against thickness is strongly modified. Especially, the sharpness of the shape resonances in the free-electron results\cite{thompsonblatt, blattthompson, paskinsingh,yu} may be lost when switching to the discrete thicknesses (it depends on the relation between the interplane separation and the oscillation period $\pi/k_F$).         

With reference to the available experimental data for Pb(111) films, we may observe that the relative amplitude of $T_c$ oscillations (related to the bulk $T_c$ value) does not excess 10~\% for the results of Guo \emph{et al.}\cite{guo} (film thickness above 22~ML), while it is even below 5~\% according to the data of Eom \emph{et al.}\cite{eom} for film thicknesses lower than 15~ML. If the comparable magnitude of oscillations might be expected for magnesium diboride, it would correspond to a few kelvins. 

The observed dampening of the short-period oscillations is partly explained by the behaviour of the matrix elements given by the formula~(\ref{eqn:matrix}), which become less dependent on the choice of the covering type when the charge spillage is allowed.

In addition, the calculated energy gap values at $T=$~0 are shown in Fig.~\ref{fig:gaps} for the thinnest films. We observe that the gap values group into two sets corresponding to $\sigma$-type and $\pi$-type subbands numbered by $n$. The behaviour of the superconducting energy gaps is consistent with the variability of $T_c$. In experimental studies of ultrathin films, we expect that only the averaged values for the two bands $\sigma$ and $\pi$ of different symmetry can be measured, while it seems unlikely to separate the individual gaps for subbands.   

\section{Summary}
In the present study, a description of BCS superconductivity in the ultrathin films was suggested, with application of the orthogonal TBA to the electron wavefunctions construction. The method allows for determination of the band structure, the Fermi level shift and finally the critical temperature of the thin superconducting films as thickness-dependent. We take into account various boundary conditions. The knowledge of the purely geometric parameters of the film is sufficient to relate the ultrathin film and the bulk matrix elements of BCS interaction. The method was applied to MgB$_2$ for which TBA seems to describe well the electronic properties. 

It was predicted that films consisting of less than 4 MgB$_2$ layers exhibit a severe decrease in the critical temperature. For the charge spillage disallowed, the critical temperature oscillates sharply with thickness (due to rapid changes of boundary conditions for different compositions of covering atomic planes) and is raised for thicker films. The charge spillage makes the dependence more smooth, damps the dependence on the kind of the covering layers and lowers in general the critical temperature below the bulk value. Such a behaviour is observed experimentally in studies of the ultrathin films of Pb(111) (see\cite{guo,eom}).    

It is usual to describe the properties of superconductors with phonon mechanism within the Eliashberg formalism, which allows for taking into account the electron-phonon interaction and phonon DOS in more accurate manner \cite{choiroundy,choi}. Especially, the distribution of the gap parameter within the distinct Fermi surface sheets was shown. The precision of the predictions based on the first-principle calculations was proven experimentally, taking the investigations of specific heat of Golubov \emph{et al.}\cite{golubov} as an example. However, we are convinced that the usage of isotropic BCS model with the parameters adjusted to fit bulk parameters of MgB$_2$ allows to capture the essentials of the ultrathin film effects in advantageously clear way and provide at least rough insight in magnesium diboride films properties. It also introduces the realistic band structure to the previously studied free-electron models\cite{thompsonblatt,blattthompson,paskinsingh}. The two-band BCS model has been succesfully applied to the thermodynamical studies of the bulk magnesium diboride by Mishonov \emph{et al.}\cite{mishonov}, who reproduced the experimental data with an accuracy of a few per cent. We believe that this model is sufficient for the purpose limited to investigating the critical temperature and energy gap values.       

\begin{acknowledgments}
The author is deeply indebted to Prof. Leszek Wojtczak for inspiration, fruitful discussions and all his helpful attitude.
\end{acknowledgments}

\end{document}